\documentclass[a4paper,11pt]{article}
\pdfoutput=1
\usepackage{jcappub} 
\usepackage{lineno}
\usepackage{array}
\usepackage{booktabs}

\newcommand{\be}{\begin{equation}}
\newcommand{\ee}{\end{equation}}
\newcommand{\ba}{\begin{eqnarray}}
\newcommand{\ea}{\end{eqnarray}}

\title{A dark energy parameterization independent constraint of the spatial  curvature $\Omega_K$ }

\author[a,b]{Zhennan Li}
\author[c,d,e]{Pengjie Zhang}

\affiliation[a]{School of Physics and Astronomy, Shanghai Jiao Tong University, Shanghai 200240, China}
\affiliation[b]{Zhiyuan College, Shanghai Jiao Tong University, Shanghai 200240, China}
\affiliation[c]{Department of Astronomy, School of Physics and Astronomy, Shanghai Jiao Tong University, Shanghai, 200240, China}
\affiliation[d]{Tsung-Dao Lee Institute, Shanghai Jiao Tong University, Shanghai
200240, China}
\affiliation[e]{Key Laboratory for Particle Astrophysics and Cosmology (MOE)/Shanghai Key Laboratory for Particle Physics and Cosmology, 
Shanghai, China}

\emailAdd{li-zhen-nan@sjtu.edu.cn; zhangpj@sjtu.edu.cn}

\abstract{Determining the spatial curvature $\Omega_K$ of the Universe has long been crucial in cosmology. In practice, this effort is often entangled with assumptions of dark energy. A combination of distance ($D_{\rm M}$, $D_{\rm L}$) and expansion rate ($H(z)$) measurements can break this degeneracy. However, fitting against discrete data points requires parameterizations of distance and expansion rate as functions of redshifts, which often induces cosmological model dependence. In this work, we propose a new dark energy model-independent parameterization of the cosmological comoving radial distance $\chi$. Fitting data combining distance ($D_{\rm M}$, $D_{\rm L}$) and Hubble parameter (or equivalently $D_H$) measurements, we are then able to obtain $\Omega_K$ in a dark energy model-independent manner. We test this parameterization and the associated fitting scheme with mock data generated with a wide range of fiducial dark energy equations of state ($-1.3<w<1.3$), finding that the best-fit $\Omega_K$ is always unbiased. Then we combine SDSS Baryon Acoustic Oscillation (BAO), Pantheon+ sample of Type Ia Supernovae (SNe Ia), and Observational Hubble Data (OHD) to constrain $\Omega_K$. We find a flat universe with $\Omega_K=-0.01\pm 0.09$. Most constraining power is contributed by SDSS BAO, with the BAO-alone constraint $\Omega_K=-0.03 \pm 0.10$. When replacing SDSS BAO with DESI year-one BAO measurement, we obtain $\Omega_K=0.06 \pm 0.08$. With the full DESI BAO data alone, we forecast $\sigma(\Omega_K)\sim 0.03$.  Our result verifies the flatness of the universe free of dark energy modeling, and the proposed parameterization would be useful for future investigation of $\Omega_K$ and other parameters of interest, such as the horizon radius. 
}

\keywords{spatial curvature; dark energy; cosmological parameters}

\begin{document}
\maketitle
\flushbottom

\section{Introduction} \label{sec:intro}

    The spatial curvature $\Omega_K$ not only influences the fate of our universe but also encodes information on the origin mechanism of the universe (e.g., \cite{Guth1981, Linde1982}),  stimulating us to conduct an accurate constraint of it. So far, most observations indicate that our universe is remarkably flat. Cosmic microwave background (CMB) experiments such as BOOMERanG \cite{Netterfield2002} and MAXIMA \cite{Hanany2000} provided the first observational proof of a flat universe \cite{de_Bernardis2000, balbi2000, Jaffe2001}. The most precise constraint on the spatial curvature $\Omega_K=-0.0001 \pm 0.0018$ was given by combining baryon acoustic oscillation (BAO) and CMB data  \cite{Alam2021, Planck2020}. Despite the observational finding of small $\Omega_K$, we can not assume a flat $\Lambda$CDM model, and further observations are needed to prove the consistency with flatness \cite{anselmi2023flat}.

    Besides, the primary CMB suffers from a well-known geometry degeneracy \cite{Bond1997, Zaldarriag1997}.  Without CMB lensing or BAO,  \cite{Planck2020} and  \cite{Planck2016b} represented a preference for closed universes with a significance slightly above 95\% confidence (the "curvature tension," \cite{Handley2021}). Independent $H_0$ measurements can also break this geometry degeneracy \cite{Planck2016b, Vagnozzi2021, Vagnozzi2021feb, Dhawan2021}.  However, given the issue of Hubble tension  (e.g., \cite{Riess2019, Riess2022}), it would be useful to constrain $\Omega_K$ independent of CMB data and local $H_0$ measurements. 

    Distance measurements, such as $D_{\rm L}$ from Type Ia supernovae (SNe Ia) and $D_{\rm M}$ from BAO, suffer from a degeneracy between $\Omega_K$ and a general dark energy model. So are $H(z)$ measurements from BAO and OHD (observational Hubble data). Combining distance measurements and $H(z)$ measurement breaks this degeneracy, allowing for dark energy model independent constraint on $\Omega_K$ through Eq.\ref{eqn:DM} \& \ref{eqn:chi} \cite{Clarkson2008}. Nevertheless, given the discrete data points and the required integration/differentiation operation, a smooth functional form of $D_{\rm M}(z)$ and $H(z)$ is required. This can be done by assuming a cosmological model such as $\Lambda$CDM with a free $\Omega_K$ or $w$CDM. This will induce model dependence in $\Omega_K$ constraint and may affect the constraint on other parameters \cite{Dossett2012}. 


    A widely adopted alternative is to parameterize the distance in a cosmological model-independent manner. 
    The most direct idea was to expand the cosmology distance as a polynomial function. However, this expansion faces a serious divergence issue in the high-redshift domain \cite{Cattoen2007}. Many new methods were proposed to overcome this problem \cite{Cattoen2007, Shafieloo2012, Aviles2014, Capozziello2020, Li2020, Zhang2023}. One of these approaches is utilizing the Padé polynomials, which successfully leads to convergence \cite{Li2020, Zhang2023}. Inspired by this method and the asymptotic behavior of the proper distance $\chi(z)$ at $z\rightarrow 0$ and $z\rightarrow \infty$,  We propose a new parameterization of $\chi(z)$. This 3-parameter parameterization has several advantages. (1) It is well behaved over all $z$, so is its derivative $d\chi/dz=c/H(z)$. (2) It provides excellent description to $w$CDM over a wide range of dark energy equation of state $w$. So we believe that it is applicable in general cases. (3) All constraining power on $\Omega_K$ then solely come from the $\chi$-$D_{\rm L,M}$ relation. Namely, the $\Omega_K$ constraint depends on no cosmological models other than the FRW metric. 

    We then apply this parameterization to the latest observations, including BAO measurements, SNe Ia, and observational Hubble data (OHD) to constrain the spatial curvature $\Omega_K$. We use the 12 data points of $D_{\rm M}/r_{\rm d}$ and $D_H/r_{\rm d}$ collected in \cite{Alam2021}, obtained from four generations of Sloan Digital Sky Survey (SDSS). This data set covers the redshift range from $z=0.38$ to $z=2.33$. For the SNe Ia data, we utilize the newest Pantheon+ sample, including 1701 light curves of 1550 unique and spectroscopically confirmed SNe Ia \cite{Pantheon+2022}. As an additional data set, we assemble 33 data points of OHD \cite{Jimenez2003, Simon2005, Stern2010, Moresco2012, Zhang2014, Moresco2015, Moresco2016, Ratsimbazafy2017, Borghi2022, jiao2023}, which provide the immediate constraint on the Hubble constant $H_0$. By applying parameter fitting to these three data sets with our model, we ought to obtain a new reliable constraint on the spatial curvature $\Omega_K$. In addition, we also apply to the BAO measurements of DESI data release 1 \cite{DESI2024III, DESI2024IV, DESI2024VI}. Our analysis confirms the flatness of the universe, independent of dark energy parameterization and CMB data. 

    This paper is organized as follows. We first introduce the method of constraining curvature in \S\ref{sec:method}, and then present the data in \S\ref{sec:data}. The results are shown and analyzed in \S\ref{sec:result}. We finally present our conclusions and discussions in \S\ref{sec:conclu}. We also include an appendix explaining the mock data test and providing further details.

\section{Method}\label{sec:method}

        We work with the following Friedmann-Robertson-Walker metric
    \be
         d\tau^2 = c^2dt^2-a^2(t)[d\chi^2+\sin^2_K(\chi)(d\theta^2+\sin^2\theta d\varphi^2)],
    \ee
    where $a(t)$ is the scale factor related to redshift as $a=(1+z)^{-1}$. $\chi$ is the comoving radial distance defined as
    \be
    \label{eqn:chi}
       \chi = c\int_0^z\frac{dz'}{H(z')}\ .
    \ee
    The special function 
    \be
       \sin_K(\chi) 
            = 
            \begin{cases}
                r_K\sin(\chi/r_K) & \Omega_K<0,\\
                \chi & \Omega_K=0,\\
                r_K\sinh(\chi/r_K) & \Omega_K>0,\\
            \end{cases}
    \ee
    where $r_K = |\Omega_K|^{-1/2}c/H_0$.

    \subsection{Breaking the curvature-dark energy degeneracy combining $H(z)$ and $D_{{\rm M,L}}$ measurements}
    It is difficult to measure $\chi(z)$ directly. Instead, we measure three main kinds of distances: the Hubble distance $D_H$, the (comoving) angular diameter distance $D_{\rm M}$, and the luminosity distance $D_{\rm L}$. 
    
    $D_H$ is directly related to the Hubble parameter $H(z)$ by
        \be
            D_H = \frac{d\chi}{dz}=\frac{c}{H(z)}\ .
        \ee
    
    The (comoving) angular diameter distance is 
        \be
        \label{eqn:DM}
            D_{\rm M} = c\sin_K\left(\chi\right).
        \ee
    Both $D_H$ and $D_{\rm M}$ can be precisely measured by BAO, up to a normalization $r_{\rm d}$, the sound horizon to the drag epoch. 
    $r_{\rm d}$ is cosmology dependent, with a value $\sim 150 $ Mpc. Without assumptions on $r_{\rm d}$ or external data set such as CMB or BBN, what BAO measures are  $D_H/r_{\rm d}$ and $D_{\rm M}/r_{\rm d}$ \cite{Alam2021}. 
    
    The luminosity distance $D_{\rm L}$ can be measured by standard candles such as Type Ia supernovae (SNe Ia) or gravitational wave standard sirens.  There exists a distance duality, 
        \be
            D_{\rm L} = D_{\rm M}(1+z).
        \ee
    For SNe Ia, what is directly measured is distance module $\mu$, defined by the difference between the apparent magnitude $m$ and the absolute magnitude $M$,
    
        \be
            \mu = m-M = 5\log_{10}\left(\frac{D_{\rm L}}{10 \rm pc}\right).
        \ee
    Besides, observational analysis of the distance module of SNe Ia requires some parameters determined by the fitting of low redshift SNe Ia data, leading to errors in the amplitude of the distance module. Therefore, we introduce a new parameter $\mu_0$ to eliminate the possible error and define a new distance module as
    \be
         \mu = 5\log_{10}D_{\rm L}+\mu_0,
    \ee
    where $\mu_0$ is a free parameter.
    
    With a single type of distance measurement, there exists a dark energy and curvature degeneracy since curvature is equivalent to a dark energy component with an equation of state $w=-1/3$. This degeneracy is automatically broken in the $w$CDM analysis where $w$ is assumed to be a constant, and in $w_0w_a$CDM where $w=w_0+w_a(1-a)$ is assumed. However, without restriction of $w$ evolution, $\Omega_K$ can not be uniquely determined. In contrast, in the combination of $D_H$ and $D_{\rm M}$ (or $D_{\rm L}$), Eq.\ref{eqn:DM} shows that $\Omega_K$ can be uniquely determined, without assumptions on $w$.

    The major motivation of our work is to measure $ \Omega_K$ independent of dark energy assumption, combining $D_H$ (and/or OHD data) and $D_{\rm M}$ (and/or $D_{\rm L}$) measurements. A further issue to overcome is that the data is only available at limited redshifts. To obtain $\chi(z)$ in Eq.\ref{eqn:DM} from $H(z)$ (or equivalently $D_H$), it is often to parameterize $H(z)$ or $\chi(z)$. For convenience of numerical calculation, we choose to parameterize $\chi(z)$ since $H(z)$ can then be obtained analytically. 
    
    \subsection{A new parameterization of $\chi(z)$}
    We propose the following three-parameter form for $\chi(z)$,
        \be
            \chi(z) = \frac{c}{H_0}\frac{z+AB[(1+z)^{3/2}-\frac{3}{2}z-1]}{1+B[(1+z)^{3/2}-1]},
        \ee
    where $A$ and $B$ are dimensionless parameters. This parameterization has the desired behavior at $z\rightarrow 0$ and $z\rightarrow \infty$. When $z\rightarrow 0$, $\chi\rightarrow cz/H_0$ and the Hubble-Lamaitre law is recovered. When $z\rightarrow \infty$, $\chi\rightarrow A(c/H_0)$. Therefore our horizon is finite, as expected from a matter-dominated universe at high redshift. 

    A further advantage of the above parameterization is that the expression for $H(z)$ is analytical. It significantly reduces the computation complexity as there is no integration. By differentiating the comoving distance, we find
    \begin{flalign}
    &H =H_0\frac{F^2(z)}{G(z)+I(z)+C}, \\ \nonumber
    &F(z) = 1+B[-1+(1+z)^{3/2}], \\ \nonumber
    &G(z) = (\frac{3}{2}AB+B-\frac{3}{2}AB^2) \sqrt{1+z},\\ \nonumber 
    &I(z) = (\frac{4}{3}AB^2-\frac{B}{2})z\sqrt{1+z}, \\ \nonumber
    &C = 1-\frac{3}{2}AB+\frac{3}{2}AB^2-B.
    \end{flalign}
    
    When $z\gg 1$, $H\propto (1+z)^{3/2}$, agreeing with the expected behavior of matter-dominated universe at $z\gg 1$. 

    We have performed various tests on this parameterization in the appendix and validated its applicability in current surveys and stage IV surveys, for $-1.3<w<0.7$. Please refer to details of these tests in the appendix. Next, we proceed to constraints on $\Omega_K$ with existing data.

\section{Data}\label{sec:data}

    \begin{table*}
    \centering
    \caption{Sloan Digital Sky Survey Baryon Acoustic Oscillation (SDSS BAO).}
    \begin{tabular*}{\textwidth}{@{\extracolsep{\fill}} l l l l l l l}
    \toprule
    $z_\mathrm{eff}$ & 0.38 & 0.51 & 0.70 & 1.48 & 2.33 & 2.33\\
    \midrule
    $D_{\rm M}/r_{\rm d}$ &$10.23\pm0.17$&$13.36\pm0.21$&$17.86\pm0.33$&$30.69\pm0.80$&$37.6\pm1.9$&$37.3\pm1.7$  \\
    \midrule
    $D_H/r_{\rm d}$ &$25.00\pm0.76$&$22.33\pm0.58$&$19.33\pm0.53$&$13.26\pm0.55$&$8.93\pm0.28$&$9.08\pm0.34$\\
    \midrule
    refer. & \cite{Alam2017}& \cite{Alam2017} & \cite{Gil2020} & \cite{Neveux2020} & \cite{Des2020} & \cite{Des2020} \\
    \bottomrule
    \end{tabular*}
    \label{tab: SDSSBAO}
    \end{table*}

    \begin{table*}
    \centering
    \caption{Dark Energy Spectroscopic Instrument year-one Baryon Acoustic Oscillation (DESI year-one BAO).}
    \begin{tabular*}{\textwidth}{@{\extracolsep{\fill}} l l l l l l }
    \toprule
    $z_\mathrm{eff}$  & 0.510 & 0.706 & 0.930 & 1.317 & 2.330\\
    \midrule
    $D_{\rm M}/r_{\rm d}$ 
    & $13.62\pm0.25$&$16.85\pm 0.32$&$21.71\pm0.28$&$27.79\pm0.69$&$39.71\pm0.94$ \\
    \midrule
    $D_H/r_{\rm d}$ &$20.98\pm 0.61$&$20.08\pm 0.60$&$17.88\pm 0.35$&$13.82\pm0.42$&$8.52\pm 0.17$\\
    \midrule
    refer. & \cite{DESI2024III}& \cite{DESI2024III} &\cite{DESI2024III} & \cite{DESI2024III} & \cite{DESI2024IV}   \\
    \bottomrule
    \end{tabular*}

    \label{tab: DESIBAO}
    \end{table*}

    \subsection{Baryon Acoustic Oscillation}
    We consider two BAO data sets. The SDSS BAO refers to \cite{Alam2021}, as shown in Table.\ref{tab: SDSSBAO}, including measurements from the SDSS-III BOSS DR12 galaxy sample\cite{Alam2017}, the SDSS-IV eBOSS DR16 luminous red galaxy (LRG) sample\cite{Gil2020}, the SDSS-VI eBOSS DR16 quasar (QSO) sample\cite{Neveux2020}, and the eBOSS DR16 autocorrelations and cross-correlations of the Ly$\alpha$ absorption and quasars\cite{Des2020}. This set of data covers 6 effective redshift points from 0.38 to 2.33.
    
    The second BAO data set is the DESI year-one BAO measurement, as shown in Table.\ref{tab: DESIBAO}, referring to \cite{DESI2024VI}. It includes BAO measurements with galaxies and quasars \cite{DESI2024III} and BAO with the Lyman-alpha forest  \cite{DESI2024IV}, which covers 5 effective redshift points from 0.510 to 2.330.

    \subsection{Type Ia Supernovae}

    We mainly refer to the latest Pantheon+ sample \cite{Pantheon+2022} for SNe Ia data. Compared with the first Pantheon release \cite{Scolnic2018}, this new sample appended 6 large samples.  Pantheon+ consists of 1701 light curves of 1550 unique, spectroscopically confirmed Type Ia supernovae across 18 different surveys \cite{Pantheon+2022}. To avoid the effect of Doppler redshift impressing the final result, we eliminate data points at redshift below $0.1$. After the selection, we make full use of 960 light curve data out of the total sample of 1701. The whole data release can be found at \url{https://pantheonplussh0es.github.io/}.

    \begin{table*}
    \centering
    \caption{Observational Hubble Data (OHD).}
    \begin{tabular*}{\textwidth}{@{\extracolsep{\fill}} l l l l l l l l}
    \toprule
    $z$ & $H(z)$ & $\sigma_H$ & refer. & $z$ & $H(z)$ & $\sigma_H$ & refer.\\
    \midrule
    0.09&69&$\pm12$&\cite{Jimenez2003} & 0.781&105&$\pm12$&\cite{Moresco2012}\\
    0.17&83&$\pm8$&\cite{Simon2005} &  0.875&125&$\pm17$&\cite{Moresco2012}\\
    0.27&77&$\pm14$&\cite{Simon2005} & 1.037&154&$\pm20$&\cite{Moresco2012}\\
    0.4&95&$\pm17$&\cite{Simon2005} & 0.07&69&$\pm19.6$&\cite{Zhang2014}\\
    0.9&117&$\pm23$&\cite{Simon2005} & 0.12&68.6&$\pm26.2$&\cite{Zhang2014}\\
    1.3&168&$\pm17$&\cite{Simon2005} & 0.20&72.9&$\pm29.6$&\cite{Zhang2014}\\
    1.43&177&$\pm18$&\cite{Simon2005} & 0.28&88.8&$\pm36.6$&\cite{Zhang2014}\\
    1.53&140&$\pm14$&\cite{Simon2005} & 1.363&160&$\pm33.6$&\cite{Moresco2015}\\
    1.75&202&$\pm40$&\cite{Simon2005} & 1.965&186.5&$\pm50.4$&\cite{Moresco2015}\\
    0.48&97&$\pm60$&\cite{Stern2010} & 0.3802&83&$\pm13.5$&\cite{Moresco2016}\\
    0.88&90&$\pm40$&\cite{Stern2010} & 0.4004&77&$\pm10.2$&\cite{Moresco2016}\\
    0.179&75&$\pm4$&\cite{Moresco2012} & 0.4247&87.1&$\pm11.2$&\cite{Moresco2016}\\
    0.199&74&$\pm5$&\cite{Moresco2012} & 0.44497&92.8&$\pm12.9$&\cite{Moresco2016}\\
    0.352&83&$\pm14$&\cite{Moresco2012} & 0.4783&80.9&$\pm9$&\cite{Moresco2016}\\
    0.593&104&$\pm13$&\cite{Moresco2012} & 0.47&89&$\pm34$&\cite{Ratsimbazafy2017}\\
    0.680&92&$\pm8$&\cite{Moresco2012} & 0.75&98.8&$\pm33.6$&\cite{Borghi2022}\\
    &&&& 0.8&  113.1& $\pm15.1$&\cite{jiao2023}\\
    \bottomrule
    \end{tabular*}
    \label{tab: OHD}
    \end{table*}

    \subsection{Observational Hubble Data}
    Determination of curvature with $D_{H,{\rm M,L}}$ is unaffected by $r_{\rm d}$ in $D_{H, {\rm M}}$ from BAO measurements and $\mu_0$ in $D_{\rm L}$ from SNe Ia measurements. However, unknown $r_{\rm d}$ and $\mu_0$ prohibit the determination of $H_0$ with BAO and SNe Ia. Just for the purpose of further check, we include the Observational Hubble Data. This enables the constraint of $H_0$ and comparison with other $H_0$ measurements, further demonstrating the applicability of our new parameterization.   This paper compiles the most recent 33 OHD points shown in Table.\ref{tab: OHD}, covering redshift from 0.09 to 1.965.

\section{Constraints on the curvature of the universe}\label{sec:result}

    \begin{table*}
    \centering
    \caption{The priors for the model parameters.}
    \begin{tabular*}{0.7\textwidth}{@{\extracolsep{\fill}} l l l }
    \toprule
    parameters & lower limit & upper limit \\
    \midrule
     $r_d$&$100$&$200$ \\
    $H_0$&$50$&$90$ \\
     A&$0$&$5$  \\
     B&$0$&$4$ \\
    $\Omega_K$&$-0.4$&$+0.4$ \\
    $\mu_0$&$20$&$30$ \\
    \bottomrule
    \end{tabular*}
    \label{tab:prior}
    \end{table*}

    We run a Markov Chain Monte Carlo (MCMC) algorithm using the emcee package \cite{Foreman2013} to fit the three sets of data. In total we have 6 free parameters, $\Omega_K$, $A$, $B$, $H_0$, $r_{\rm d}$ and $\mu_0$. We adopt uniform priors on these parameters within the corresponding ranges listed in Table.\ref{tab:prior}. 
    
\subsection{SDSS BAO + Pantheon+ + OHD}

\begin{figure*}
		\centering
 		\includegraphics[width=0.8\textwidth]{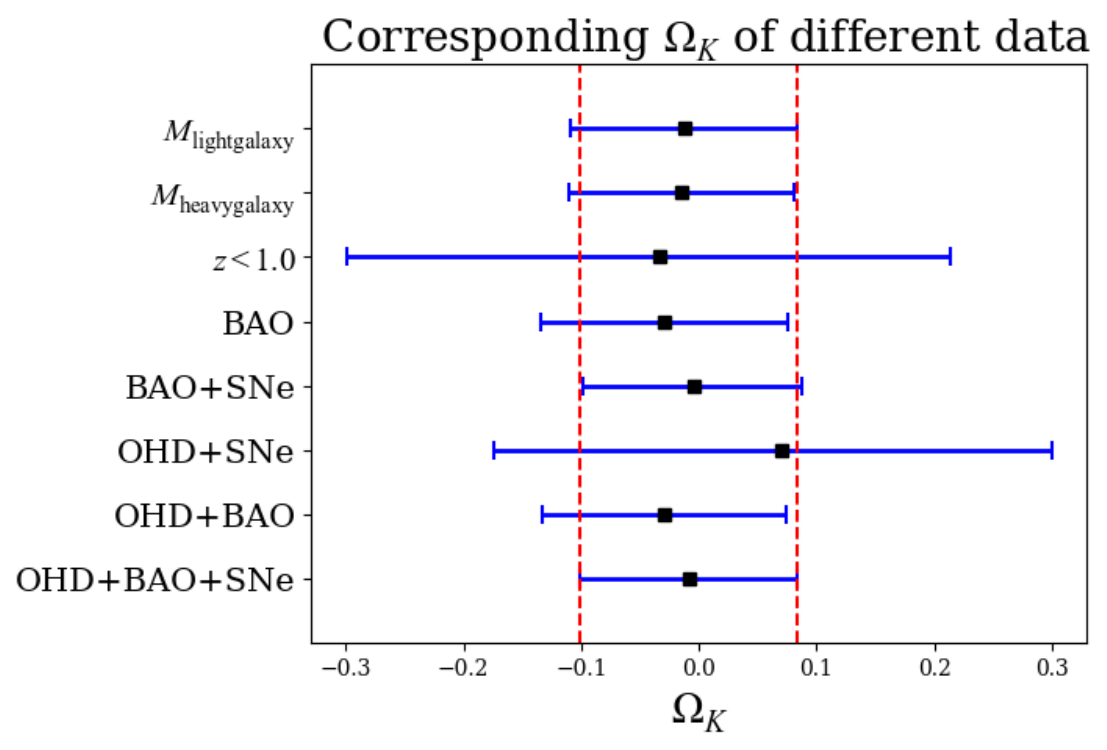}
 		\caption{The mean value of $\Omega_K$ and $1\sigma$ given by fitting with different kind of data combinations. Here, we use SDSS BAO. BAO plays the most crucial role in constraining curvature, while SNe Ia drives it from negative to positive.  \label{fig:curve}}
    \end{figure*}

    Figure.\ref{fig:curve} shows the constraint on $\Omega_K$. 
    \be
        \Omega_K=-0.01\pm 0.09\ , \ {\rm SDSS\ BAO+ {Pantheon+} +OHD}.
    \ee
    This is consistent with a flat universe. Although the error is relatively large, it is a constraint independent of CMB and independent of dark energy modeling. 
    
\begin{figure*}
		\centering
 		\includegraphics[width=\textwidth]{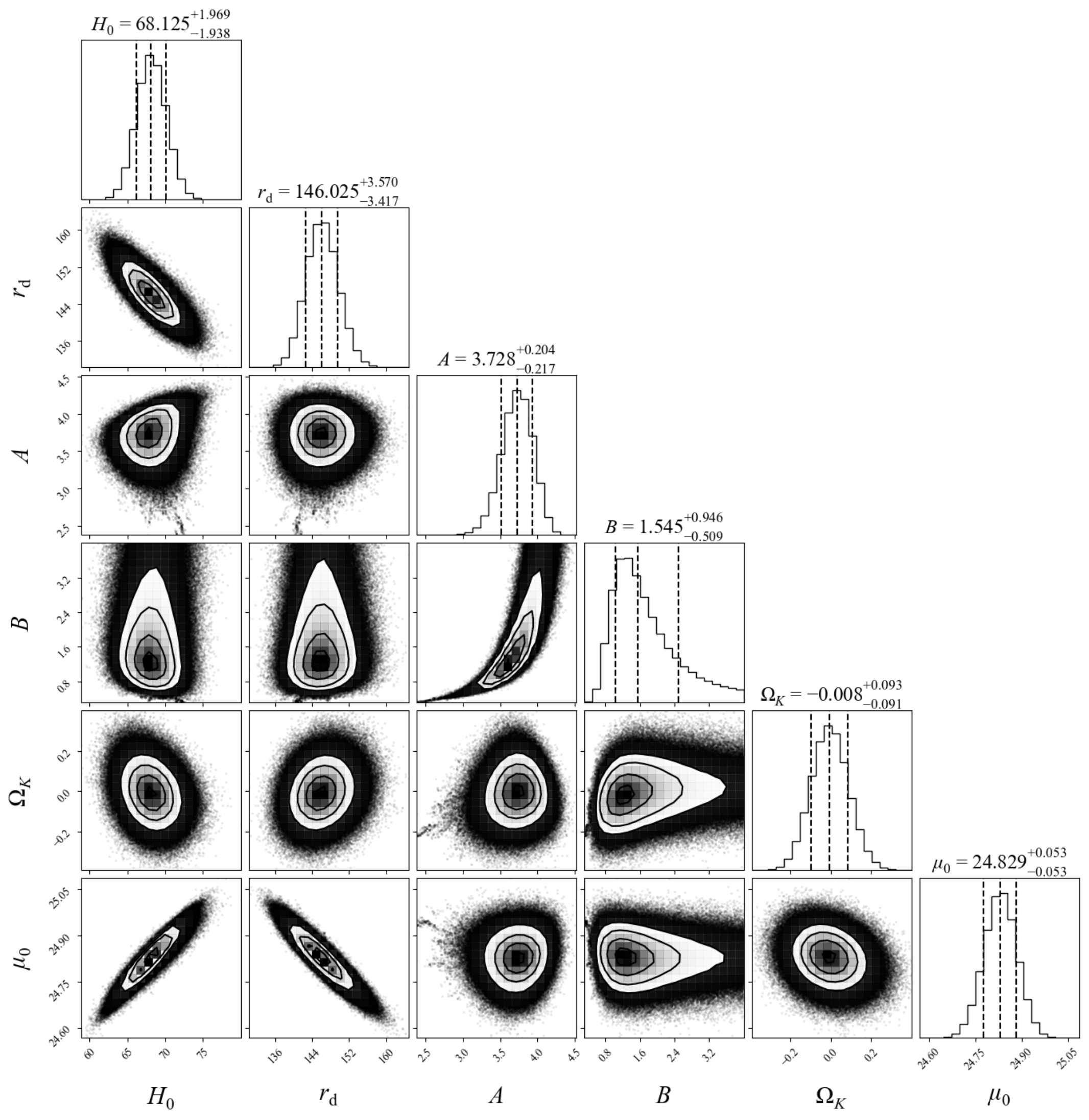}
 		\caption{MCMC results of the data fitting using our model, combining SDSS BAO, Patheon+ SNe Ia, and OHD. The best-fit values of each parameter are displayed on the top of the corresponding column.}
            \label{fig:his}
    \end{figure*}

    Figure.\ref{fig:his}  shows the posterior distributions of all 6 parameters of our model, using SDSS BAO. Constraint on $H_0$ also agrees with the Planck results, further demonstrating the applicability of our method. 

\begin{figure*}
		\centering
 		\includegraphics[width=1\textwidth]{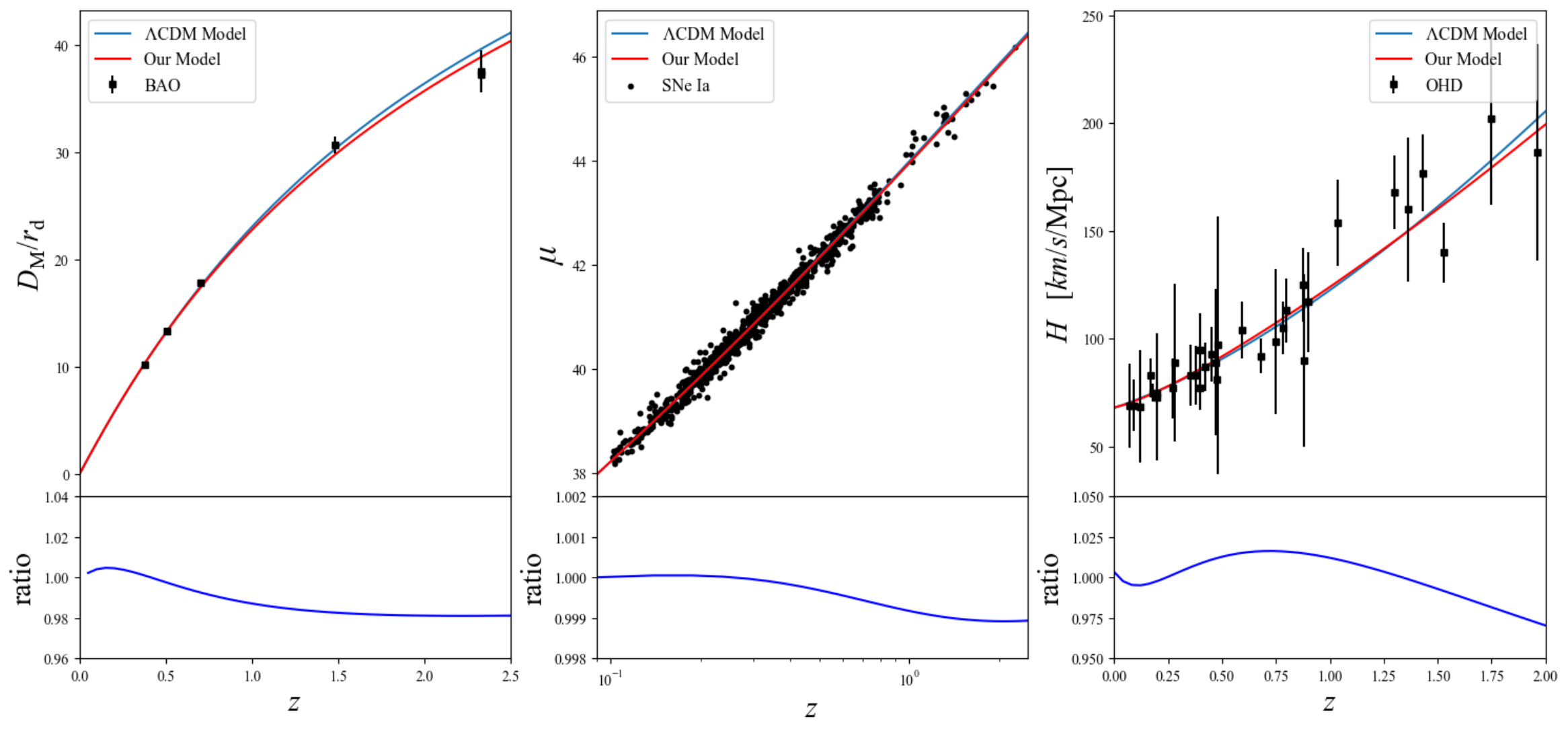}
 		\caption{Final fitting result of angular diameter distance $D_{\rm M}$ over sound horizon $r_{\rm d}$, absolute magnitude $\mu$, and Hubble parameter $H$. The black points are SDSS BAO, Patheon+ SNe Ia, and OHD. The blue line is the best fitting curve of the $\Lambda$CDM model, and the red line is the fitting curve using our model. The bottom panels show the ratio of the two best fitting results, from left to right, defined as $H_{\rm our}/H_{\Lambda {\rm CDM}}$, $(D_M/r_{\rm d})_{\rm our}/(D_M/r_{\rm d})_{\Lambda {\rm CDM}}$, and $\mu_{\rm our}/\mu_{\Lambda {\rm CDM}}$.\label{fig:H}}
    \end{figure*}

    Figure.\ref{fig:H} represents $H(z)$, $D_{\rm M}(z)$ and $\mu(z)$  predicted by the best-fitting results of our model. They agree with the data excellently. We also show the best-fitting results under the $\Lambda$CDM cosmology. The two differ at the level of ${\mathcal O}(1\%)$, however, the differences are too small to be distinguished by the given data sets.

    We conduct a test to show how much contribution is given by each data set. From Figure.\ref{fig:curve}, we can find that BAO plays the most crucial role in constraining curvature, while SNe Ia tends to drive it from negative to positive. The OHD contribution to the $\Omega_K$ constraint is negligible, while its contribution to $H_0$ is significant.  We also check the contribution from different redshifts. 
    Since the impact of curvature only shows up when $\chi$ is sufficiently large ($\chi\sim c/H_0$), the constraint from the $z<1.0$ data points is significantly weaker than the full data. 

    We also perform one consistency test. For SNe Ia, the constraint on $\Omega_K$ should not be related to host galaxy properties such as the stellar mass. To test it, we split SNe Ia into two subsamples of low and high stellar mass\footnote{we first divide the SNe Ia sample into eight bins by redshift, then in each bin, we divide the SNe Ia into two sets of equal numbers according to their host galaxy's stellar mass. We then form two subsamples of SNe Ia. }. From Figure.\ref{fig:curve}, we see that the $\Omega_K$ constraint is nearly unchanged.

\subsection{DESI year-one BAO + Pantheon+ + OHD}

    \begin{figure*}
		\centering
 		\includegraphics[width=1\textwidth]{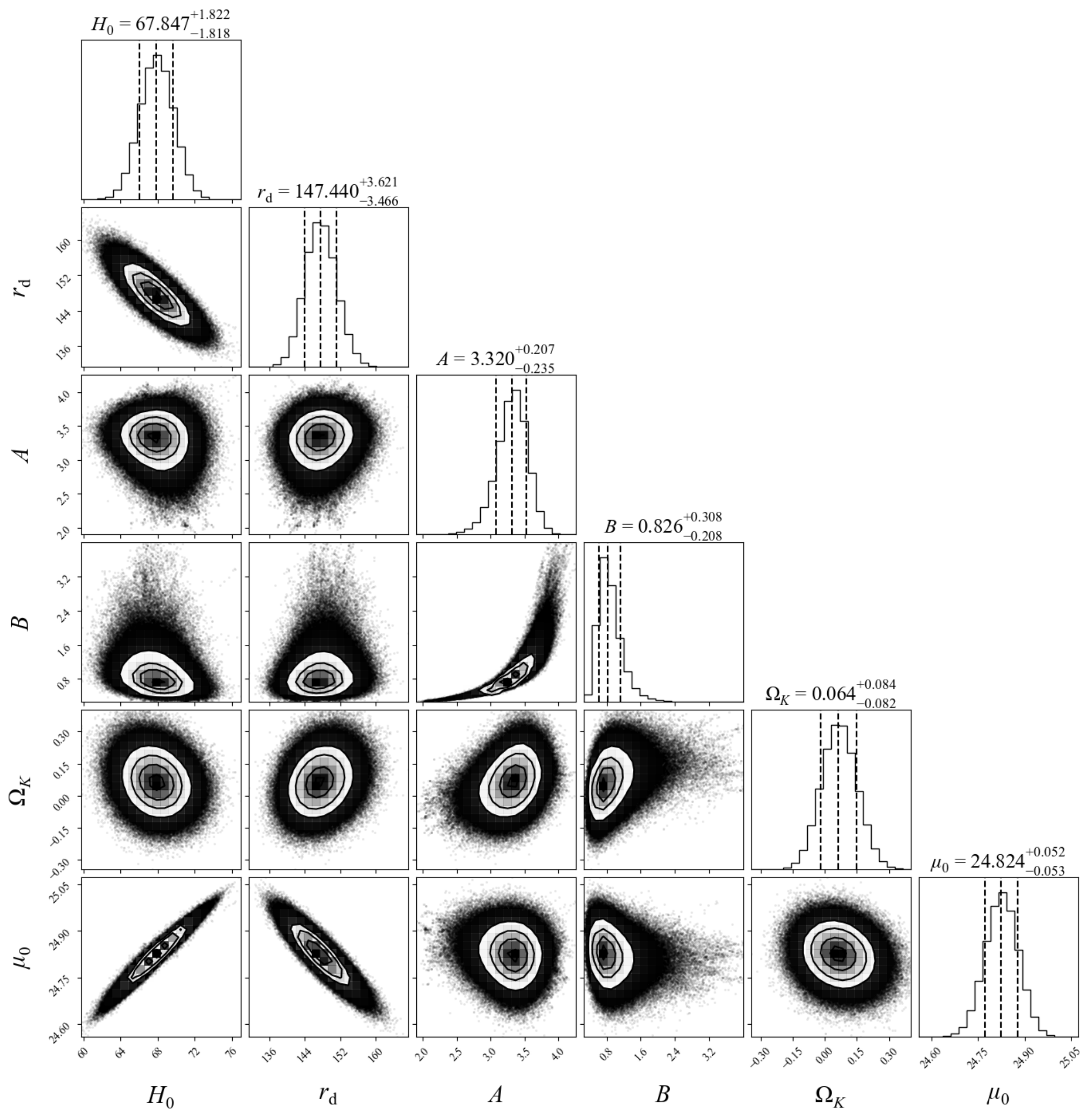}
 		\caption{MCMC results of the data fitting using our model, combining DESI year-one BAO, Patheon+ SNe Ia, and OHD. The best-fit values of each parameter are displayed on the top of the corresponding column.}
            \label{fig:DESIhis}
    \end{figure*}

    \begin{figure*}
		\centering
 		\includegraphics[width=1\textwidth]{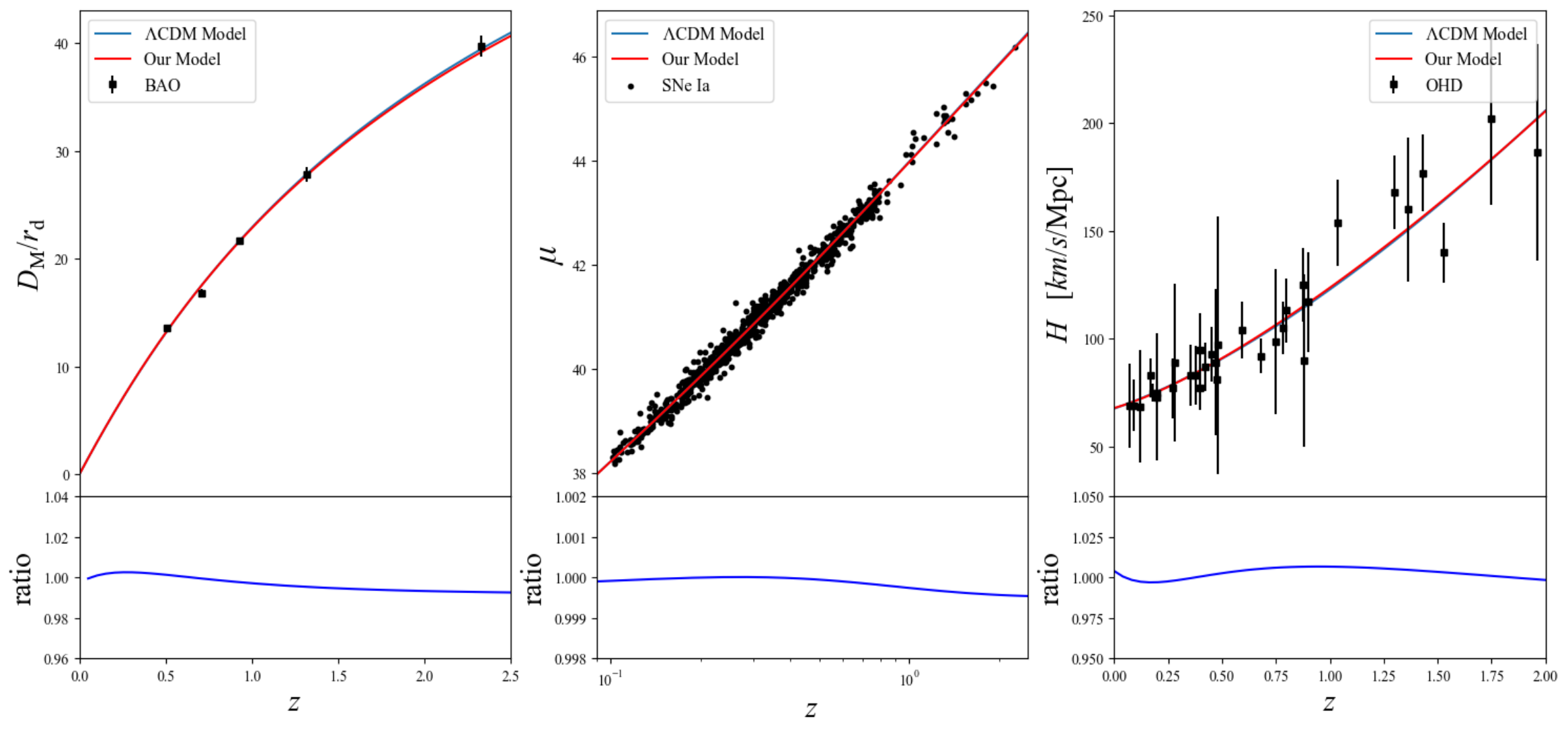}
 		\caption{Final fitting result of angular diameter distance $D_{\rm M}$ over sound horizon $r_{\rm d}$, absolute magnitude $\mu$, and Hubble parameter $H$. The black points with the error bar are DESI year-one BAO, Patheon+ SNe Ia, and OHD. The blue line is the best fitting curve of the $\Lambda$CDM model, and the red line is the fitting curve using our model. The bottom panels show the ratio of the two best fitting results, from left to right, defined as $H_{\rm our}/H_{\Lambda \rm{CDM}}$, $(D_{\rm M}/r_{\rm d})_{\rm our}/(D_{\rm M}/r_{\rm d})_{\Lambda {\rm CDM}}$, and $\mu_{\rm our}/\mu_{\Lambda {\rm CDM}}$.\label{fig:HDESI}}
    \end{figure*}

    We also update SDSS BAO with DESI year-one BAO measurements and obtain
    \be
     \Omega_K=0.06\pm 0.08\ ,\ {\rm DESI\  BAO+{Pantheon+}+OHD}\ .
    \ee
    The DESI collaboration constrained $\Omega_K=0.065^{+0.068}_{-0.078}$ assuming the $\Lambda$CDM+$\Omega_K$ cosmology, and $\Omega_K=0.085^{+0.100}_{-0.085}$ assuming the $w_0w_a$CDM+$\Omega_K$ cosmology (Table 3 in \cite{DESI2024VI}). These constraints are consistent with our results. Figure.\ref{fig:HDESI} shows the curves of $H(z)$, $D_{\rm M}(z)$ and $\mu(z)$ predicted by the best-fit values of our method, along with that predicted by the best-fit $\Lambda$CDM.  Interestingly, the two agree at the level much better than $1\%$. This is different from the case of SDSS BAO data. This is a demonstration of the discrepancy between the SDSS BAO measurement and the DESI BAO measurement.  

    As shown earlier, the major contribution to the $\Omega_K$ constraint is BAO. The full DESI data will significantly improve the BAO measurement and therefore the $\Omega_K$ constraint. Given the DESI BAO forecast \cite{DESI2024}, our method is expected to constrain $\Omega_K$ independent of dark energy modeling, with an error (Figure.\ref{fig:OmegaK}) 
    \be
        \sigma(\Omega_K)\simeq 0.03\ ,\ {\rm full\ DESI\ BAO}\ .
    \ee

\section{Conclusion}\label{sec:conclu}

    In this paper, we propose a three-parameter parameterization of the comoving radial distance $\chi$ as a function of $z$, validated to describe $\Lambda$CDM and $w$CDM with excellent accuracy. Applying this parameterization to data of BAO, SNe Ia, and OHD, we can constrain $\Omega_K$ through the $\chi$-$D_{\rm M, L}$ relation. We obtain $\Omega_K=-0.01 \pm 0.09$ with SDSS BAO and $\Omega_K=0.06 \pm 0.08$ after replacing SDSS BAO with DESI BAO. BAO mainly contributes to the constraint. SDSS BAO alone constrains $\Omega_K=-0.03\pm 0.10$, DESI year-one BAO alone constrains $\Omega_K=0.09\pm 0.09$, which is consistent with $\Omega_K = 0.135\pm0.087$ in \cite{Jiang2024}. Using our model, We expect the full DESI BAO to achieve $\sigma(\Omega_K)=0.03$.  Our result verifies the universe's flatness free of dark energy modeling, and the proposed parameterization would be useful for future investigation of this issue and other parameters of interest, like assuming a non-standard sound horizon to constrain spatial curvature \cite{Stevens2023}.

\section{acknowledgments}
    This work was supported by the National Key R\&D Program of China (2023YFA1607800, 2023YFA1607801, 2020YFC2201602), the China Manned Space Project (\#CMS-CSST-2021-A02), and the Fundamental Research Funds for the Central Universities. This work uses numpy \cite{Harris2020}, matplotlib \cite{Hunter2007}, scipy \cite{Virtanen2020}, corner \cite{2016JOSS....1...24F}, and  emcee \cite{Foreman2013}.

\appendix

\section{Mock Data}\label{sec:mock}

    \begin{figure*}
        \centering
        \includegraphics[width=0.8\textwidth]{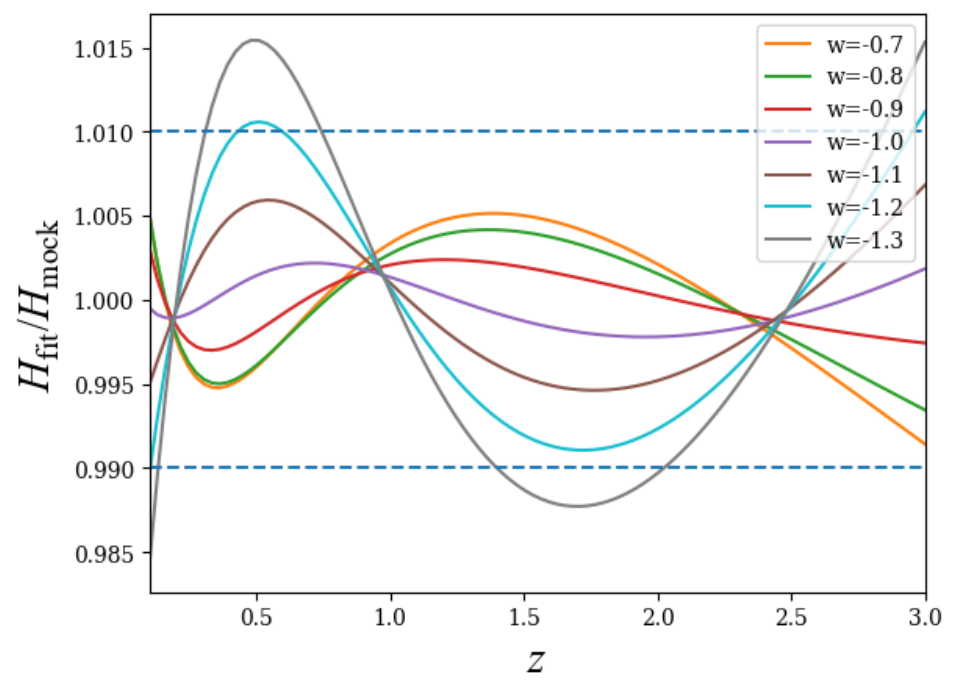}
        \caption{{\textbf {Mock 1}}: The percentage form of fitting result. Here $H_{\rm fit}$ is the best fitting function of our model and $H_{\rm mock}$ is the mock function generated referring to DESI forecasts. In the most fitting range, our results deviate from the mock data no more than 1\% marked by the blue dashed line. \label{fig:Hp1}}
    \end{figure*}

    \begin{figure*}
        \centering
        \includegraphics[width=0.8\textwidth]{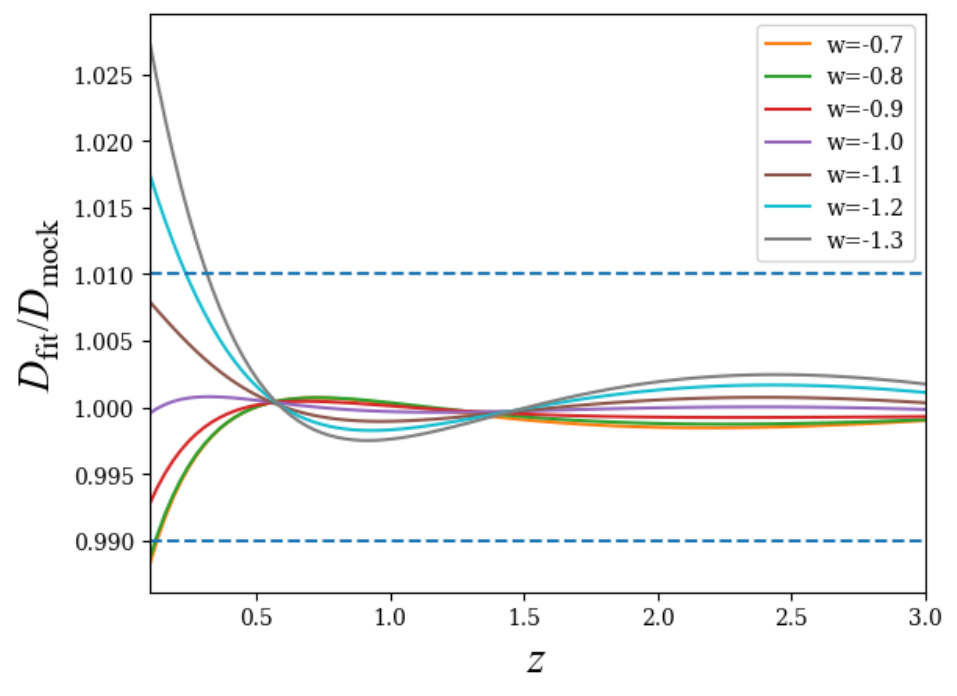}
        \caption{\textbf{Mock 1}: The percentage form of fitting result. Here, $D_{\rm fit}$ is the best fitting function, and   $D_{\rm mock}$ is the mock function generated referring to DESI forecasts. In the most fitting range, our results deviate from the mock data significantly smaller than 1\% marked by the blue dashed line.  \label{fig:Dp1}}
    \end{figure*}

    Here we validate the proposed parameterization against two sets of mock data. Both mocks adopt $w$CDM with curvature as the fiducial cosmology in which
    
    \be \label{eq.wcdm}                     
        H=H_0\sqrt{\Omega_m(1+z)^3+\Omega_K(1+z)^2+\Omega_\Lambda(1+z)^{3(1+w)}}.  
    \ee
    where $w$ is the equation of state of the dark energy.  We fix $H_0 = 67.4 km/s/{\rm Mpc}$ and $\Omega_m = 0.3153$. We consider $w\in[-1.3,-0.7]$ and several values of $\Omega_K$.The mock data points are generated using the fiducial cosmology prediction with two sets of errors, one corresponding to the full DESI BAO measurements, and one corresponding to the current data. We denote them as mock 1 and mock 2. 
    
    \subsection{Mock 1: full DESI BAO forecast}

    Mock 1 adopts the redshifts and errors represented in the DESI forecasts \cite{DESI2024}, with 35 data points ($z\in [0.05,3.45]$) in total.  We adopt $\Omega_K=-0.00096$ (and therefore $\Omega_\Lambda=0.6847$).  Figure.\ref{fig:Hp1} \& \ref{fig:Dp1} show $Hr_{\rm d}$ and $D/r_{\rm d}$ predicted by the best-fit parameters using our model. It agrees with the fiducial curves to better than $\sim 1\%$ accuracy for $w\in[-1.2,-0.7]$. Interestingly, the agreement is better than $0.5\%$ for the $\Lambda$CDM cosmology (purple curve).  Furthermore, the constrained $\Omega_K$ is always unbiased, for a wide range of $\Omega_K\in [-0.1,0.1]$ and $w\in [-1.3,-0.7]$. These results demonstrate that our model is applicable to future data such as DESI BAO.  This also implies that our model is applicable to current data. Nevertheless, we create the other mock (Mock 2) to explicitly validate it.

    \begin{figure*}
        \centering
        \includegraphics[width=0.8\textwidth]{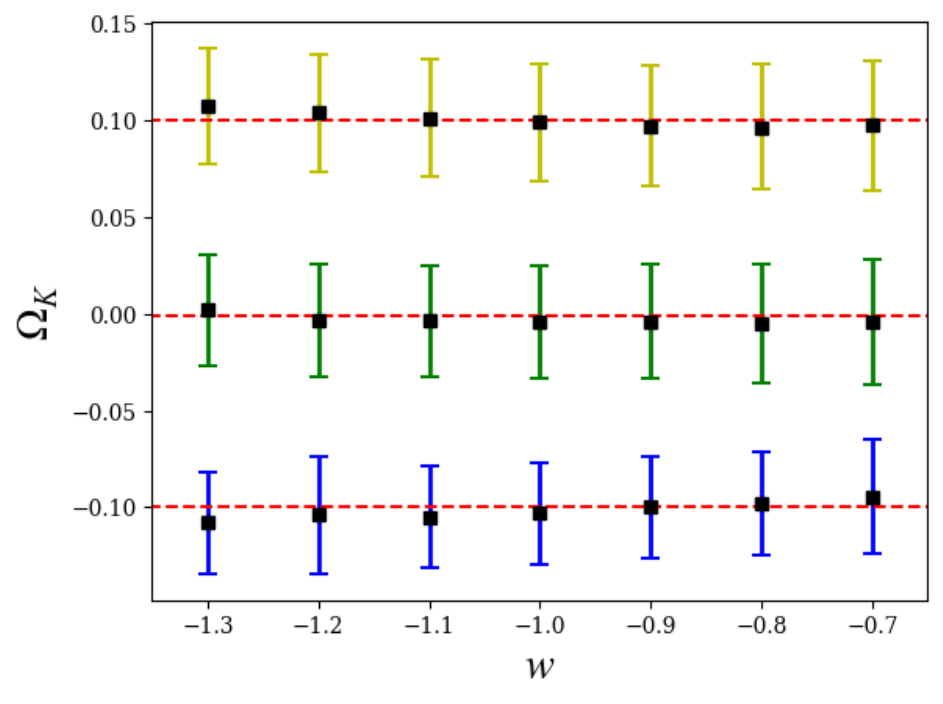}
        \caption{\textbf{Mock 1}: The fitting result of $\Omega_K$ with fiducial model of different equation of state $w$. All the results agree with the truth values whether it is a universe with positive curvature, flat curvature, or negative curvature. From top to bottom, the three red dashed lines correspond to $\Omega_K =0.10$, $\Omega_K = -0.00096$, and $\Omega_K = -0.10$ respectively.\label{fig:OmegaK}}
    \end{figure*}

    \subsection{Mock 2: current data used in this paper}

    Mock 2 just replaces the central values of SDSS BAO/Pantheon+ SNe Ia/OHD data with that of the fiducial cosmology while keeping the error bars unchanged. We adopt $r_{\rm d} =146 {\rm Mpc}$ and $\mu_0 = 25$. Changing these nuisance parameters has no impact on cosmological constraints.  Table.\ref{tab:mock} shows the best-fit values, which agree excellently with the input values. In particular, $\Omega_K$ is unbiased within the error bar. To better demonstrate that our model/parameterization can describe the true $H(z)$ ($D(z)$ and $\mu(z)$ excellently, we define the agreement using an effective $\chi^2$, 
    \be
     \chi^2_f = \sum \frac{(f_{fitting} - f_{fiducial})^2}{\sigma^2_f}
    \ee
    Here $f$ denotes $H$ for OHD, $D_{{\rm M},H}$ for BAO and $\mu$ for SNe Ia, respectively. Notice that the mock data centers at the fiducial cosmology prediction, so we do not expect $\chi^2/$d.o.f.$\sim 1$. Instead, $\chi^2\ll 1$ is expected if our model well describes the mock data. 
    This testing promises the validity of utilizing our novel model to make the data fit with the data listed in \S\ref{sec:data}. Table. \ref{tab:chi2} shows that this is indeed the case for all types of data. Therefore, our model is applicable to current data.

    \begin{table*}
    \centering
    \caption{\textbf{Mock 2}: The best fitting and $68.3\%$ confidence levels of the parameters in our model compared with their truth values used in generating mock data under $w$CDM model.}
    \begin{tabular*}{\textwidth}{@{\extracolsep{\fill}} l l l l l l l}
    \toprule
     $\quad$ & $r_{\rm d}$ &$H_0$& A& B & $\Omega_K$ & $\mu_0$\\
    \midrule
   fiducial model &146&67.4&/&/&-0.00096&25\\
    \midrule w=-0.7&$146.6^{+3.5}_{-3.4}$&$67.6^{+1.9}_{-1.9}$&$3.43^{+0.15}_{-0.16}$&$1.51^{+0.64}_{-0.39}$&$0.018_{-0.111}^{+0.114}$&$25.00_{-0.05}^{+0.05}$  \\
    \midrule
    w=-0.9&$145.5^{+3.6}_{-3.5}$&$67.7^{+1.9}_{-1.9}$&$3.36^{+0.22}_{-0.25}$&$0.92^{+0.39}_{-0.25}$&$0.011_{-0.100}^{+0.102}$&$25.00_{-0.05}^{+0.05}$  \\
    \midrule
    w=-1.1&$146.1^{+3.7}_{-3.6}$&$67.1^{+1.8}_{-1.8}$&$2.91^{+0.35}_{-0.47}$&$0.49^{+0.19}_{-0.14}$&$0.011_{-0.092}^{+0.094}$&$25.00_{-0.06}^{+0.05}$  \\
    \midrule    w=-1.3&$145.8^{+3.8}_{-3.6}$&$65.8^{+1.7}_{-1.7}$&$2.25^{+0.51}_{-0.74}$&$0.31^{+0.12}_{-0.09}$&$0.009_{-0.085}^{+0.087}$&$25.00_{-0.06}^{+0.06}$  \\          
    \bottomrule
    \end{tabular*}
    \label{tab:mock}
    \end{table*}

    \begin{table*}
    \centering
    \caption{\textbf{Mock 2}: Final $\chi^2$ result for each data set and added up $\chi^2$ using mock data generated in the second method.}
    \begin{tabular*}{\textwidth}{@{\extracolsep{\fill}} l l l l l l }
    \toprule
    $\quad$ &  $\chi^2_H$ & $\chi^2_{D_H}$& $\chi^2_{D_{\rm M}}$ & $\chi^2_{\mu}$ & $\chi^2_{total}$    \\
    \midrule
    w=-0.7&$0.0604$&$0.4989$&$0.0333$&$0.0787$&$0.6713$  \\
    \midrule
    w=-0.9&$0.0562$&$0.3356$&$0.0351$&$0.1338$&$0.5606$  \\
    \midrule
     w=-1.1&$0.0359$&$0.1314$&$0.0412$&$0.2371$&$0.4456$  \\
    \midrule
    w=-1.3&$0.0144$&$0.1416$&$0.0240$&$0.0661$&$0.2461$  \\
    \bottomrule
    \end{tabular*}
    \label{tab:chi2}
    \end{table*}





\bibliographystyle{JHEP}
\bibliography{main.bib}

\providecommand{\href}[2]{#2}\begingroup\raggedright\begin{thebibliography}{10}

\bibitem{Guth1981}
A.H.~{Guth}, \emph{{Inflationary universe: A possible solution to the horizon
  and flatness problems}},
  \href{https://doi.org/10.1103/PhysRevD.23.347}{\emph{Physical Review D}
  {\bfseries 23} (1981) 347}.

\bibitem{Linde1982}
A.D.~{Linde}, \emph{{A new inflationary universe scenario: A possible solution
  of the horizon, flatness, homogeneity, isotropy and primordial monopole
  problems}}, \href{https://doi.org/10.1016/0370-2693(82)91219-9}{\emph{Physics
  Letters B} {\bfseries 108} (1982) 389}.

\bibitem{Netterfield2002}
C.B.~Netterfield, P.A.R.~Ade, J.J.~Bock, J.R.~Bond, J.~Borrill, A.~Boscaleri
  et~al., \emph{A measurement by boomerang of multiple peaks in the angular
  power spectrum of the cosmic microwave background},
  \href{https://doi.org/10.1086/340118}{\emph{The Astrophysical Journal}
  {\bfseries 571} (2002) 604–614}.

\bibitem{Hanany2000}
S.~Hanany, P.~Ade, A.~Balbi, J.~Bock, J.~Borrill, A.~Boscaleri et~al.,
  \emph{Maxima-1: A measurement of the cosmic microwave background anisotropy
  on angular scales of 10[arcmin]–5°},
  \href{https://doi.org/10.1086/317322}{\emph{The Astrophysical Journal}
  {\bfseries 545} (2000) L5–L9}.

\bibitem{de_Bernardis2000}
P.~de~Bernardis, P.A.R.~Ade, J.J.~Bock, J.R.~Bond, J.~Borrill, A.~Boscaleri
  et~al., \emph{A flat universe from high-resolution maps of the cosmic
  microwave background radiation},
  \href{https://doi.org/10.1038/35010035}{\emph{Nature} {\bfseries 404} (2000)
  955–959}.

\bibitem{balbi2000}
A.~{Balbi}, P.~{Ade}, J.~{Bock}, J.~{Borrill}, A.~{Boscaleri}, P.~{De
  Bernardis} et~al., \emph{{Erratum: Constraints on Cosmological Parameters
  from MAXIMA-1}}, \href{https://doi.org/10.1086/323608}{\emph{The
  Astrophysical Journal Letters} {\bfseries 558} (2001) L145}
  [\href{https://arxiv.org/abs/astro-ph/0005124}{{\ttfamily
  astro-ph/0005124}}].

\bibitem{Jaffe2001}
A.H.~Jaffe, P.A.R.~Ade, A.~Balbi, J.J.~Bock, J.R.~Bond, J.~Borrill et~al.,
  \emph{Cosmology from maxima-1, boomerang, and cobe dmr cosmic microwave
  background observations},
  \href{https://doi.org/10.1103/PhysRevLett.86.3475}{\emph{Physical Review
  Letters} {\bfseries 86} (2001) 3475}.

\bibitem{Alam2021}
S.~{Alam}, M.~{Aubert}, S.~{Avila}, C.~{Balland}, J.E.~{Bautista},
  M.A.~{Bershady} et~al., \emph{{Completed SDSS-IV extended Baryon Oscillation
  Spectroscopic Survey: Cosmological implications from two decades of
  spectroscopic surveys at the Apache Point Observatory}},
  \href{https://doi.org/10.1103/PhysRevD.103.083533}{\emph{Physical Review D}
  {\bfseries 103} (2021) 083533}
  [\href{https://arxiv.org/abs/2007.08991}{{\ttfamily 2007.08991}}].

\bibitem{Planck2020}
{Planck Collaboration}, N.~{Aghanim}, Y.~{Akrami}, M.~{Ashdown}, J.~{Aumont},
  C.~{Baccigalupi} et~al., \emph{{Planck 2018 results. VI. Cosmological
  parameters}},
  \href{https://doi.org/10.1051/0004-6361/201833910}{\emph{Astronomy \&
  Astrophysics} {\bfseries 641} (2020) A6}
  [\href{https://arxiv.org/abs/1807.06209}{{\ttfamily 1807.06209}}].

\bibitem{anselmi2023flat}
S.~Anselmi, M.F.~Carney, J.T.~Giblin, S.~Kumar, J.B.~Mertens, M.~O'Dwyer
  et~al., \emph{What is flat $\lambda$cdm, and may we choose it?},
  {\emph{Journal of Cosmology and Astroparticle Physics} {\bfseries 2023}
  (2023) 049}.

\bibitem{Bond1997}
J.R.~{Bond}, G.~{Efstathiou} and M.~{Tegmark}, \emph{{Forecasting cosmic
  parameter errors from microwave background anisotropy experiments}},
  \href{https://doi.org/10.1093/mnras/291.1.L33}{\emph{Monthly Notices of the
  Royal Astronomical Society} {\bfseries 291} (1997) L33}
  [\href{https://arxiv.org/abs/astro-ph/9702100}{{\ttfamily
  astro-ph/9702100}}].

\bibitem{Zaldarriag1997}
M.~{Zaldarriaga}, D.N.~{Spergel} and U.~{Seljak}, \emph{{Microwave Background
  Constraints on Cosmological Parameters}},
  \href{https://doi.org/10.1086/304692}{\emph{The Astrophysical Journal}
  {\bfseries 488} (1997) 1}
  [\href{https://arxiv.org/abs/astro-ph/9702157}{{\ttfamily
  astro-ph/9702157}}].

\bibitem{Planck2016b}
{Planck Collaboration}, P.A.R.~{Ade}, N.~{Aghanim}, M.~{Arnaud}, M.~{Ashdown},
  J.~{Aumont} et~al., \emph{{Planck 2015 results. XIII. Cosmological
  parameters}},
  \href{https://doi.org/10.1051/0004-6361/201525830}{\emph{Astronomy \&
  Astrophysics} {\bfseries 594} (2016) A13}
  [\href{https://arxiv.org/abs/1502.01589}{{\ttfamily 1502.01589}}].

\bibitem{Handley2021}
W.~{Handley}, \emph{{Curvature tension: Evidence for a closed universe}},
  \href{https://doi.org/10.1103/PhysRevD.103.L041301}{\emph{Physical Review D}
  {\bfseries 103} (2021) L041301}
  [\href{https://arxiv.org/abs/1908.09139}{{\ttfamily 1908.09139}}].

\bibitem{Vagnozzi2021}
S.~{Vagnozzi}, E.~{Di Valentino}, S.~{Gariazzo}, A.~{Melchiorri}, O.~{Mena} and
  J.~{Silk}, \emph{{The galaxy power spectrum take on spatial curvature and
  cosmic concordance}},
  \href{https://doi.org/10.1016/j.dark.2021.100851}{\emph{Physics of the Dark
  Universe} {\bfseries 33} (2021) 100851}
  [\href{https://arxiv.org/abs/2010.02230}{{\ttfamily 2010.02230}}].

\bibitem{Vagnozzi2021feb}
S.~{Vagnozzi}, A.~{Loeb} and M.~{Moresco}, \emph{{Eppur {\`e} piatto? The
  Cosmic Chronometers Take on Spatial Curvature and Cosmic Concordance}},
  \href{https://doi.org/10.3847/1538-4357/abd4df}{\emph{The Astrophysical
  Journal} {\bfseries 908} (2021) 84}
  [\href{https://arxiv.org/abs/2011.11645}{{\ttfamily 2011.11645}}].

\bibitem{Dhawan2021}
S.~{Dhawan}, J.~{Alsing} and S.~{Vagnozzi}, \emph{{Non-parametric spatial
  curvature inference using late-Universe cosmological probes}},
  \href{https://doi.org/10.1093/mnrasl/slab058}{\emph{Monthly Notices of the
  Royal Astronomical Society} {\bfseries 506} (2021) L1}
  [\href{https://arxiv.org/abs/2104.02485}{{\ttfamily 2104.02485}}].

\bibitem{Riess2019}
A.G.~{Riess}, S.~{Casertano}, W.~{Yuan}, L.M.~{Macri} and D.~{Scolnic},
  \emph{{Large Magellanic Cloud Cepheid Standards Provide a 1\% Foundation for
  the Determination of the Hubble Constant and Stronger Evidence for Physics
  beyond {\ensuremath{\Lambda}}CDM}},
  \href{https://doi.org/10.3847/1538-4357/ab1422}{\emph{The Astrophysical
  Journal} {\bfseries 876} (2019) 85}
  [\href{https://arxiv.org/abs/1903.07603}{{\ttfamily 1903.07603}}].

\bibitem{Riess2022}
A.G.~{Riess}, W.~{Yuan}, L.M.~{Macri}, D.~{Scolnic}, D.~{Brout}, S.~{Casertano}
  et~al., \emph{{A Comprehensive Measurement of the Local Value of the Hubble
  Constant with 1 km s$^{-1}$ Mpc$^{-1}$ Uncertainty from the Hubble Space
  Telescope and the SH0ES Team}},
  \href{https://doi.org/10.3847/2041-8213/ac5c5b}{\emph{The Astrophysical
  Journal Letters} {\bfseries 934} (2022) L7}
  [\href{https://arxiv.org/abs/2112.04510}{{\ttfamily 2112.04510}}].

\bibitem{Clarkson2008}
C.~{Clarkson}, B.~{Bassett} and T.H.-C.~{Lu}, \emph{{A General Test of the
  Copernican Principle}},
  \href{https://doi.org/10.1103/PhysRevLett.101.011301}{\emph{Physical Review
  Letters} {\bfseries 101} (2008) 011301}
  [\href{https://arxiv.org/abs/0712.3457}{{\ttfamily 0712.3457}}].

\bibitem{Dossett2012}
J.N.~{Dossett} and M.~{Ishak}, \emph{{Spatial curvature and cosmological tests
  of general relativity}},
  \href{https://doi.org/10.1103/PhysRevD.86.103008}{\emph{Physical Review D}
  {\bfseries 86} (2012) 103008}
  [\href{https://arxiv.org/abs/1205.2422}{{\ttfamily 1205.2422}}].

\bibitem{Cattoen2007}
C.~{Catto{\"e}n} and M.~{Visser}, \emph{{The Hubble series: convergence
  properties and redshift variables}},
  \href{https://doi.org/10.1088/0264-9381/24/23/018}{\emph{Classical and
  Quantum Gravity} {\bfseries 24} (2007) 5985}
  [\href{https://arxiv.org/abs/0710.1887}{{\ttfamily 0710.1887}}].

\bibitem{Shafieloo2012}
A.~{Shafieloo}, \emph{{Crossing statistic: reconstructing the expansion history
  of the universe}},
  \href{https://doi.org/10.1088/1475-7516/2012/08/002}{\emph{Journal of
  Cosmology and Astroparticle Physics} {\bfseries 2012} (2012) 002}
  [\href{https://arxiv.org/abs/1204.1109}{{\ttfamily 1204.1109}}].

\bibitem{Aviles2014}
A.~{Aviles}, C.~{Gruber}, O.~{Luongo} and H.~{Quevedo}, \emph{{Constraints from
  Cosmography in various parameterizations}},
  \href{https://doi.org/10.48550/arXiv.1301.4044}{\emph{arXiv e-prints} (2013)
  arXiv:1301.4044} [\href{https://arxiv.org/abs/1301.4044}{{\ttfamily
  1301.4044}}].

\bibitem{Capozziello2020}
S.~{Capozziello}, R.~{D'Agostino} and O.~{Luongo}, \emph{{High-redshift
  cosmography: auxiliary variables versus Pad{\'e} polynomials}},
  \href{https://doi.org/10.1093/mnras/staa871}{\emph{Monthly Notices of the
  Royal Astronomical Society} {\bfseries 494} (2020) 2576}
  [\href{https://arxiv.org/abs/2003.09341}{{\ttfamily 2003.09341}}].

\bibitem{Li2020}
E.-K.~{Li}, M.~{Du} and L.~{Xu}, \emph{{General cosmography model with spatial
  curvature}}, \href{https://doi.org/10.1093/mnras/stz3308}{\emph{Monthly
  Notices of the Royal Astronomical Society} {\bfseries 491} (2020) 4960}
  [\href{https://arxiv.org/abs/1903.11433}{{\ttfamily 1903.11433}}].

\bibitem{Zhang2023}
K.~{Zhang}, T.~{Zhou}, B.~{Xu}, Q.~{Huang} and Y.~{Yuan}, \emph{{Joint
  Constraints on the Hubble Constant, Spatial Curvature, and Sound Horizon from
  the Late-time Universe with Cosmography}},
  \href{https://doi.org/10.3847/1538-4357/acee6e}{\emph{The Astrophysical
  Journal} {\bfseries 957} (2023) 5}
  [\href{https://arxiv.org/abs/2310.16512}{{\ttfamily 2310.16512}}].

\bibitem{Pantheon+2022}
D.~{Scolnic}, D.~{Brout}, A.~{Carr}, A.G.~{Riess}, T.M.~{Davis}, A.~{Dwomoh}
  et~al., \emph{{The Pantheon+ Analysis: The Full Data Set and Light-curve
  Release}}, \href{https://doi.org/10.3847/1538-4357/ac8b7a}{\emph{The
  Astrophysical Journal} {\bfseries 938} (2022) 113}
  [\href{https://arxiv.org/abs/2112.03863}{{\ttfamily 2112.03863}}].

\bibitem{Jimenez2003}
R.~{Jimenez}, L.~{Verde}, T.~{Treu} and D.~{Stern}, \emph{{Constraints on the
  Equation of State of Dark Energy and the Hubble Constant from Stellar Ages
  and the Cosmic Microwave Background}},
  \href{https://doi.org/10.1086/376595}{\emph{The Astrophysical Journal}
  {\bfseries 593} (2003) 622}
  [\href{https://arxiv.org/abs/astro-ph/0302560}{{\ttfamily
  astro-ph/0302560}}].

\bibitem{Simon2005}
J.~{Simon}, L.~{Verde} and R.~{Jimenez}, \emph{{Constraints on the redshift
  dependence of the dark energy potential}},
  \href{https://doi.org/10.1103/PhysRevD.71.123001}{\emph{Physical Review D}
  {\bfseries 71} (2005) 123001}
  [\href{https://arxiv.org/abs/astro-ph/0412269}{{\ttfamily
  astro-ph/0412269}}].

\bibitem{Stern2010}
D.~{Stern}, R.~{Jimenez}, L.~{Verde}, M.~{Kamionkowski} and S.A.~{Stanford},
  \emph{{Cosmic chronometers: constraining the equation of state of dark
  energy. I: H(z) measurements}},
  \href{https://doi.org/10.1088/1475-7516/2010/02/008}{\emph{Journal of
  Cosmology and Astroparticle Physics} {\bfseries 2010} (2010) 008}
  [\href{https://arxiv.org/abs/0907.3149}{{\ttfamily 0907.3149}}].

\bibitem{Moresco2012}
M.~{Moresco}, L.~{Verde}, L.~{Pozzetti}, R.~{Jimenez} and A.~{Cimatti},
  \emph{{New constraints on cosmological parameters and neutrino properties
  using the expansion rate of the Universe to z
  \raisebox{-0.5ex}\textasciitilde 1.75}},
  \href{https://doi.org/10.1088/1475-7516/2012/07/053}{\emph{Journal of
  Cosmology and Astroparticle Physics} {\bfseries 2012} (2012) 053}
  [\href{https://arxiv.org/abs/1201.6658}{{\ttfamily 1201.6658}}].

\bibitem{Zhang2014}
C.~{Zhang}, H.~{Zhang}, S.~{Yuan}, S.~{Liu}, T.-J.~{Zhang} and Y.-C.~{Sun},
  \emph{{Four new observational H(z) data from luminous red galaxies in the
  Sloan Digital Sky Survey data release seven}},
  \href{https://doi.org/10.1088/1674-4527/14/10/002}{\emph{Research in
  Astronomy and Astrophysics} {\bfseries 14} (2014) 1221}
  [\href{https://arxiv.org/abs/1207.4541}{{\ttfamily 1207.4541}}].

\bibitem{Moresco2015}
M.~{Moresco}, \emph{{Raising the bar: new constraints on the Hubble parameter
  with cosmic chronometers at z \raisebox{-0.5ex}\textasciitilde 2.}},
  \href{https://doi.org/10.1093/mnrasl/slv037}{\emph{Monthly Notices of the
  Royal Astronomical Society} {\bfseries 450} (2015) L16}
  [\href{https://arxiv.org/abs/1503.01116}{{\ttfamily 1503.01116}}].

\bibitem{Moresco2016}
M.~{Moresco}, L.~{Pozzetti}, A.~{Cimatti}, R.~{Jimenez}, C.~{Maraston},
  L.~{Verde} et~al., \emph{{A 6\% measurement of the Hubble parameter at
  z\raisebox{-0.5ex}\textasciitilde0.45: direct evidence of the epoch of cosmic
  re-acceleration}},
  \href{https://doi.org/10.1088/1475-7516/2016/05/014}{\emph{Journal of
  Cosmology and Astroparticle Physics} {\bfseries 2016} (2016) 014}
  [\href{https://arxiv.org/abs/1601.01701}{{\ttfamily 1601.01701}}].

\bibitem{Ratsimbazafy2017}
A.L.~{Ratsimbazafy}, S.I.~{Loubser}, S.M.~{Crawford}, C.M.~{Cress},
  B.A.~{Bassett}, R.C.~{Nichol} et~al., \emph{{Age-dating luminous red galaxies
  observed with the Southern African Large Telescope}},
  \href{https://doi.org/10.1093/mnras/stx301}{\emph{Monthly Notices of the
  Royal Astronomical Society} {\bfseries 467} (2017) 3239}
  [\href{https://arxiv.org/abs/1702.00418}{{\ttfamily 1702.00418}}].

\bibitem{Borghi2022}
N.~{Borghi}, M.~{Moresco} and A.~{Cimatti}, \emph{{Toward a Better
  Understanding of Cosmic Chronometers: A New Measurement of H(z) at z 0.7}},
  \href{https://doi.org/10.3847/2041-8213/ac3fb2}{\emph{The Astrophysical
  Journal Letters} {\bfseries 928} (2022) L4}
  [\href{https://arxiv.org/abs/2110.04304}{{\ttfamily 2110.04304}}].

\bibitem{jiao2023}
K.~{Jiao}, N.~{Borghi}, M.~{Moresco} and T.-J.~{Zhang}, \emph{{New
  Observational H(z) Data from Full-spectrum Fitting of Cosmic Chronometers in
  the LEGA-C Survey}},
  \href{https://doi.org/10.3847/1538-4365/acbc77}{\emph{The Astrophysical
  Journal Supplement Series} {\bfseries 265} (2023) 48}
  [\href{https://arxiv.org/abs/2205.05701}{{\ttfamily 2205.05701}}].

\bibitem{DESI2024III}
{DESI Collaboration}, A.G.~{Adame}, J.~{Aguilar}, S.~{Ahlen}, S.~{Alam},
  D.M.~{Alexander} et~al., \emph{{DESI 2024 III: Baryon Acoustic Oscillations
  from Galaxies and Quasars}},
  \href{https://doi.org/10.48550/arXiv.2404.03000}{\emph{arXiv e-prints} (2024)
  arXiv:2404.03000} [\href{https://arxiv.org/abs/2404.03000}{{\ttfamily
  2404.03000}}].

\bibitem{DESI2024IV}
{DESI Collaboration}, A.G.~{Adame}, J.~{Aguilar}, S.~{Ahlen}, S.~{Alam},
  D.M.~{Alexander} et~al., \emph{{DESI 2024 IV: Baryon Acoustic Oscillations
  from the Lyman Alpha Forest}},
  \href{https://doi.org/10.48550/arXiv.2404.03001}{\emph{arXiv e-prints} (2024)
  arXiv:2404.03001} [\href{https://arxiv.org/abs/2404.03001}{{\ttfamily
  2404.03001}}].

\bibitem{DESI2024VI}
{DESI Collaboration}, A.G.~{Adame}, J.~{Aguilar}, S.~{Ahlen}, S.~{Alam},
  D.M.~{Alexander} et~al., \emph{{DESI 2024 VI: Cosmological Constraints from
  the Measurements of Baryon Acoustic Oscillations}},
  \href{https://doi.org/10.48550/arXiv.2404.03002}{\emph{arXiv e-prints} (2024)
  arXiv:2404.03002} [\href{https://arxiv.org/abs/2404.03002}{{\ttfamily
  2404.03002}}].

\bibitem{Alam2017}
S.~{Alam}, M.~{Ata}, S.~{Bailey}, F.~{Beutler}, D.~{Bizyaev}, J.A.~{Blazek}
  et~al., \emph{{The clustering of galaxies in the completed SDSS-III Baryon
  Oscillation Spectroscopic Survey: cosmological analysis of the DR12 galaxy
  sample}}, \href{https://doi.org/10.1093/mnras/stx721}{\emph{Monthly Notices
  of the Royal Astronomical Society} {\bfseries 470} (2017) 2617}
  [\href{https://arxiv.org/abs/1607.03155}{{\ttfamily 1607.03155}}].

\bibitem{Gil2020}
H.~{Gil-Mar{\'\i}n}, J.E.~{Bautista}, R.~{Paviot}, M.~{Vargas-Maga{\~n}a},
  S.~{de la Torre}, S.~{Fromenteau} et~al., \emph{{The Completed SDSS-IV
  extended Baryon Oscillation Spectroscopic Survey: measurement of the BAO and
  growth rate of structure of the luminous red galaxy sample from the
  anisotropic power spectrum between redshifts 0.6 and 1.0}},
  \href{https://doi.org/10.1093/mnras/staa2455}{\emph{Monthly Notices of the
  Royal Astronomical Society} {\bfseries 498} (2020) 2492}
  [\href{https://arxiv.org/abs/2007.08994}{{\ttfamily 2007.08994}}].

\bibitem{Neveux2020}
R.~{Neveux}, E.~{Burtin}, A.~{de Mattia}, A.~{Smith}, A.J.~{Ross}, J.~{Hou}
  et~al., \emph{{The completed SDSS-IV extended Baryon Oscillation
  Spectroscopic Survey: BAO and RSD measurements from the anisotropic power
  spectrum of the quasar sample between redshift 0.8 and 2.2}},
  \href{https://doi.org/10.1093/mnras/staa2780}{\emph{Monthly Notices of the
  Royal Astronomical Society} {\bfseries 499} (2020) 210}
  [\href{https://arxiv.org/abs/2007.08999}{{\ttfamily 2007.08999}}].

\bibitem{Des2020}
H.~{du Mas des Bourboux}, J.~{Rich}, A.~{Font-Ribera}, V.~{de Sainte Agathe},
  J.~{Farr}, T.~{Etourneau} et~al., \emph{{The Completed SDSS-IV Extended
  Baryon Oscillation Spectroscopic Survey: Baryon Acoustic Oscillations with
  Ly{\ensuremath{\alpha}} Forests}},
  \href{https://doi.org/10.3847/1538-4357/abb085}{\emph{The Astrophysical
  Journal} {\bfseries 901} (2020) 153}
  [\href{https://arxiv.org/abs/2007.08995}{{\ttfamily 2007.08995}}].

\bibitem{Scolnic2018}
D.M.~{Scolnic}, D.O.~{Jones}, A.~{Rest}, Y.C.~{Pan}, R.~{Chornock},
  R.J.~{Foley} et~al., \emph{{The Complete Light-curve Sample of
  Spectroscopically Confirmed SNe Ia from Pan-STARRS1 and Cosmological
  Constraints from the Combined Pantheon Sample}},
  \href{https://doi.org/10.3847/1538-4357/aab9bb}{\emph{The Astrophysical
  Journal} {\bfseries 859} (2018) 101}
  [\href{https://arxiv.org/abs/1710.00845}{{\ttfamily 1710.00845}}].

\bibitem{Foreman2013}
D.~{Foreman-Mackey}, D.W.~{Hogg}, D.~{Lang} and J.~{Goodman}, \emph{{emcee: The
  MCMC Hammer}}, \href{https://doi.org/10.1086/670067}{\emph{Publications of
  the Astronomical Society of the Pacific} {\bfseries 125} (2013) 306}
  [\href{https://arxiv.org/abs/1202.3665}{{\ttfamily 1202.3665}}].

\bibitem{DESI2024}
{DESI Collaboration}, A.G.~{Adame}, J.~{Aguilar}, S.~{Ahlen}, S.~{Alam},
  G.~{Aldering} et~al., \emph{{Validation of the Scientific Program for the
  Dark Energy Spectroscopic Instrument}},
  \href{https://doi.org/10.3847/1538-3881/ad0b08}{\emph{The Astronomical
  Journal} {\bfseries 167} (2024) 62}
  [\href{https://arxiv.org/abs/2306.06307}{{\ttfamily 2306.06307}}].

\bibitem{Jiang2024}
J.-Q.~{Jiang}, D.~{Pedrotti}, S.S.~{da Costa} and S.~{Vagnozzi},
  \emph{{Nonparametric late-time expansion history reconstruction and
  implications for the Hubble tension in light of recent DESI and type Ia
  supernovae data}},
  \href{https://doi.org/10.1103/PhysRevD.110.123519}{\emph{Physical Review D}
  {\bfseries 110} (2024) 123519}
  [\href{https://arxiv.org/abs/2408.02365}{{\ttfamily 2408.02365}}].

\bibitem{Stevens2023}
J.~{Stevens}, H.~{Khoraminezhad} and S.~{Saito}, \emph{{Constraining the
  spatial curvature with cosmic expansion history in a cosmological model with
  a non-standard sound horizon}},
  \href{https://doi.org/10.1088/1475-7516/2023/07/046}{\emph{Journal of
  Cosmology and Astroparticle Physics} {\bfseries 2023} (2023) 046}
  [\href{https://arxiv.org/abs/2212.09804}{{\ttfamily 2212.09804}}].

\bibitem{Harris2020}
C.R.~{Harris}, K.J.~{Millman}, S.J.~{van der Walt}, R.~{Gommers},
  P.~{Virtanen}, D.~{Cournapeau} et~al., \emph{{Array programming with NumPy}},
  \href{https://doi.org/10.1038/s41586-020-2649-2}{\emph{Nature} {\bfseries
  585} (2020) 357} [\href{https://arxiv.org/abs/2006.10256}{{\ttfamily
  2006.10256}}].

\bibitem{Hunter2007}
J.D.~{Hunter}, \emph{{Matplotlib: A 2D Graphics Environment}},
  \href{https://doi.org/10.1109/MCSE.2007.55}{\emph{Computing in Science and
  Engineering} {\bfseries 9} (2007) 90}.

\bibitem{Virtanen2020}
P.~{Virtanen}, R.~{Gommers}, T.E.~{Oliphant}, M.~{Haberland}, T.~{Reddy},
  D.~{Cournapeau} et~al., \emph{{SciPy 1.0: fundamental algorithms for
  scientific computing in Python}},
  \href{https://doi.org/10.1038/s41592-019-0686-2}{\emph{Nature Methods}
  {\bfseries 17} (2020) 261}
  [\href{https://arxiv.org/abs/1907.10121}{{\ttfamily 1907.10121}}].

\bibitem{2016JOSS....1...24F}
D.~{Foreman-Mackey}, \emph{{corner.py: Scatterplot matrices in Python}},
  \href{https://doi.org/10.21105/joss.00024}{\emph{The Journal of Open Source
  Software} {\bfseries 1} (2016) 24}.

\end{thebibliography}\endgroup

\end{document}